\documentclass[fleqn,twoside]{article}
\usepackage{espcrc2}

\usepackage{epsfig}
\usepackage[figuresright]{rotating}

\newcommand{\ga}{\gamma}

\newcommand{\ep}{\epsilon}

\newcommand{\beq}{\begin{equation}}
\newcommand{\eeq}{\end{equation}}
\newcommand{\bdm}{\begin{displaymath}}
\newcommand{\edm}{\end{displaymath}}
\newcommand{\bea}{\begin{eqnarray}}
\newcommand{\eea}{\end{eqnarray}}
\newcommand{\<}{\langle}
\renewcommand{\>}{\rangle}
\newcommand{\ovr}{\over}

\renewcommand{\dag}{^\dagger}

\newcommand{\qbar}{\bar q}
\newcommand{\ubar}{\bar u}
\newcommand{\dbar}{\bar d}
\newcommand{\mr}{\mathrm}

\newcommand{\ms}{m_\mr{sea}}
\newcommand{\mv}{m_\mr{val}}
\newcommand{\Nf}{N_{\!f\,}}%{N_\mr{f\,}}
\newcommand{\Mpi}{M_{\pi}}
\newcommand{\Fpi}{F_{\pi}}
\newcommand{\MeV}{\,\mr{MeV}}
\newcommand{\GeV}{\,\mr{GeV}}

\newcommand{\fm}{\,\mr{fm}}
\newcommand{\MSbar}{{\overline{\mr{MS}}}}

\title{\vspace{-14mm}%
{\normalsize DESY 02-160\hfill{\tt hep-ph/0209319}}\\[-2mm]
{\normalsize September 2002}\\[4mm]
Lattice QCD data versus Chiral Perturbation Theory: the case of $M_\pi$}

\author{S. D\"urr\address[DESY]{DESY, Platanenallee 6, 15738 Zeuthen, Germany}}

\begin{document}

\begin{abstract}
I present a selection of recent lattice data by major collaborations for the
pseudo-Goldstone boson masses in full ($N_\mr{f}\!=\!2$) QCD, where the valence
quarks are chosen exactly degenerate with the sea quarks. At least the more
chiral points should be consistent with Chiral Perturbation Theory for the
latter to be useful in extrapolating to physical masses. Perspectives to
reliably determine NLO Gasser-Leutwyler coefficients are discussed.
\vspace{-0.1pc}
\end{abstract}

\maketitle

%%%%%%%%%%%%%%%%%%%%%%%%%%%%%%%%%%%%%%%%%%%%%%%%%%%%%%%%%%%%%%%%%%%%%%%%%%%%%%%

\section{INTRODUCTION}

In lattice QCD physical observables like masses and decay constants are
extracted from the fall-off pattern of correlation functions; for instance
\bea
C(x)\!&\!=\!&\!\<Q(x)\dag Q(0)\>
\nonumber
\\
{  }\!&\!=\!&\!{1\ovr Z}\int\!DU D\qbar Dq \; e^{-S_g-S_f} \; Q(x)\dag Q(0)
\label{corrone}
\eea
with $Q(x)=\ubar(x) \Gamma d(x)$ and
$\Gamma\in\{\ga_{{}_5},\ga_{{}_4}\ga_{{}_5}\}$
allows to determine $\Mpi$ and $\Fpi$.
Like in the continuum, the fermions are integrated out and (\ref{corrone})
reads
\bea
\<\dbar(x) \Gamma u(x) \!\!\!&\!\!\!\!&\!\!\!\ubar(0) \Gamma d(0)\>=
\nonumber
\\
\!\!\!&\!\!\!\!&\!\!\!{1\ovr Z}\int\!DU \det(D\!+\!\ms) e^{-S_g} \times 
\\
\!\!\!&\!\!\!\!&\!\!\!\mr{Tr}\;
\Gamma(D\!+\!\mv)^{-1}_{x0}\Gamma(D\!+\!\mv)^{-1}_{0x}
\nonumber
\eea
where $\mr{Tr}$ goes over colour and spinor indices.
In this form, the quark masses $\ms, \mv$ should be equal, since they stem from
the very same term $S_f\!=\!\int\!dx\,\qbar(D\!+\!m)q$ in the original form,
eqn.~(\ref{corrone}).

However, state-of-the-art algorithms for the determinant $\det(D\!+\!m)$ and
the Green's function $S(x,0)\!=\!((D\!+\!\mv)^{-1})_{x0}$ get parametrically
slow as the quark mass is taken light.
Since this effect is much more pronounced for the determinant, most groups
are restricted to a regime where the {\em sea\/}-quark mass for the two light
flavours (the one in the determinant) is about half of the physical strange
mass, while the {\em valence\/}-quark mass for the $u/d$-quarks (the one
in the propagator) may be pushed down to $\sim\!1/4$ of the physical $m_s$.

With these data at hand, one has to {\em extrapolate\/} both in $\ms$ and $\mv$
down to the physical $u$- and $d$-masses (which we take degenerate); see Fig\,1.
Since the extrapolation range is large, it is crucial to know the functional
form against which the data may be matched.
Partially quenched chiral perturbation theory (PQ-XPT) provides such a
framework \cite{PQ-XPT}, but its limited range of applicability (around the
twofold chiral limit) might cause a problem; it is a priori not clear whether
the bulk of the existing data lies in this range.

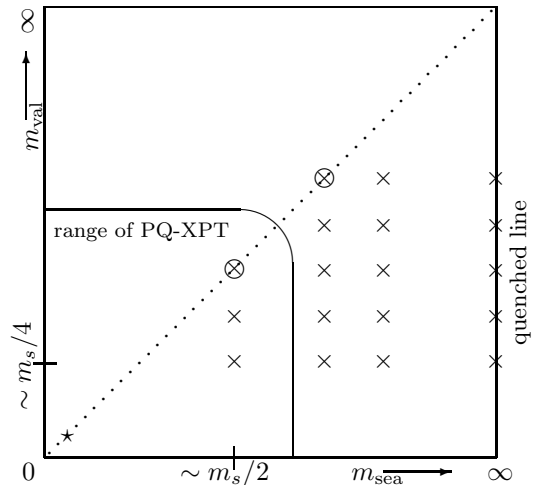
\begin{figure}[!b]%1
\vspace{-7mm}%\vspace{-6mm}
\begin{center}
\unitlength 0.6cm
\begin{picture}(12,11)(0,0)
\linethickness{0.2mm}
\put(01,01){\line(1,0){10}}
\put(11,01){\line(0,1){10}}
\put(01,11){\line(1,0){10}}
\put(01,01){\line(0,1){10}}
\put(.5,.5){$0$}
\put(5.2,0.75){\line(0,1){0.5}}
\put(0.75,3.1){\line(1,0){0.5}}
\put(4,0.5){$\sim m_s/2$}
\put(7.8,0.5){$m_\mr{sea}$}
\put(8.5,0.7){\vector(1,0){1.5}}
\put(10.8,.5){$\infty$}
\put(0.25,2.05){\begin{turn}{90}{$\sim m_s/4$}\end{turn}}
\put(0.6,7.8){\begin{turn}{90}{$m_\mr{val}$}\end{turn}}
\put(0.0,8.5){\begin{turn}{90}{\vector(1,0){1.5}}\end{turn}}
\put(0.5,10.4){\begin{turn}{90}{$\infty$}\end{turn}}
\multiput(1.1,1.1)(0.2,0.2){50}{\circle*{0.05}}
\put(4.975,3){$\times$}
\put(4.975,4){$\times$}
\put(4.975,5.050){$\times$}
\put(6.975,3){$\times$}
\put(6.975,4){$\times$}
\put(6.975,5){$\times$}
\put(6.975,6){$\times$}
\put(6.975,7.050){$\times$}
\put(8.275,3){$\times$}
\put(8.275,4){$\times$}
\put(8.275,5){$\times$}
\put(8.275,6){$\times$}
\put(8.275,7.050){$\times$}
\put(10.77,3){$\times$}
\put(10.77,4){$\times$}
\put(10.77,5){$\times$}
\put(10.77,6){$\times$}
\put(10.77,7.050){$\times$}
\put(1.355,1.355){$\star$}
\put(5.2,5.2){\circle{0.4}}
\put(7.2,7.2){\circle{0.4}}
\put(11.3,3.5){\begin{turn}{90}{\small quenched line}\end{turn}}
\put(1,1){\oval(11,11)[tr]}
\put(1.2,5.9){\footnotesize range of PQ-XPT}
\end{picture}
\end{center}
\vspace{-16mm}
\caption{\small Schematic view of partially quenched data taking, with
crosses indicating measurements in the $(m_\mr{sea},m_\mr{val})$ plane for
degenerate u/d-quarks. While the real-world pion ($\star$) is on the unitary
line (diagonal), substantial savings in CPU-time stem from taking the detour
through the sector with $m_\mr{val}\!<\!m_\mr{sea}$.}
\end{figure}

%%%%%%%%%%%%%%%%%%%%%%%%%%%%%%%%%%%%%%%%%%%%%%%%%%%%%%%%%%%%%%%%%%%%%%%%%%%%%%%

\section{ELEMENTS OF XPT}

The mandatory test is most conveniently done along the diagonal line in Fig.\,1,
which is the domain of (ordinary) XPT \cite{Gasser:1983yg,Gasser:1984gg}.
This is the effective theory of QCD at energies small compared to the
intrinsic scale, $4\pi\Fpi$, which faithfully reproduces the structure in
the chiral limit.
Its Green's functions are dominated by pseudo-Goldstone contributions and
computed in a systematic expansion in the external momenta and the quark mass,
the latter counted as $m\!\sim\!p^2$ \cite{Gasser:1983yg}.

The leading part of the Lagrangian, $L^{(2)}$, contains two low-energy
constants ($B, F$) which are finite dimensionful quantities%
\footnote{$F$ is the pion decay constant in the chiral limit $\sim\!87\MeV$,
while $B\!=\!-\Sigma/F^2$ with $\Sigma$ the chiral (one-flavour) condensate
(in the 2- or 3-flavour theory) in the chiral limit.} and the quark mass
matrix $M\!=\!\mr{diag}(m_d,m_u\,[,m_s])$.
The $O(p^2)$ suppressed part, $L^{(4)}$, contains about a dozen so-called
Gasser-Leutwyler (GL) coefficients, denoted $l_i$ for 2 flavours
\cite{Gasser:1983yg} and $L_i$ if the strange  quark is included
\cite{Gasser:1984gg}.
In the dimensionally regulated theory, the $l_i$ or $L_i$ consist of a
divergent part ($\propto\!\ep^{-1}$, where $\ep\!=\!4\!-\!d$) and a finite part
($\propto\!\ep^0$) which is scale-dependent.
After some reshuffling of finite contributions in $\MSbar$ style, the latter
part (denoted by $l_i^\mr{r}(\mu)$ or $L_i^\mr{r}(\mu)$) encodes for the QCD
short distance contributions.
In an arbitrary Green's function calculated at order $O(p^4)$, the divergence
from the loop compensates exactly the divergent part of the GL-coefficient (see
Fig.~2), and an important check of the finite part is that the scale-dependent
(finite) integral $\bar I(\mu)$ and the renormalized ``constants''
$l_i^\mr{r}(\mu)$ (or $L_i^\mr{r}(\mu)$) add up to form an exactly
{\em scale-independent\/} expression.

\begin{figure}[!b]%2
\vspace{-18mm}
\begin{center}
\unitlength 0.55cm
\begin{picture}(12,4)(0,0)
\linethickness{0.2mm}
\put(1,1){$\Sigma_\pi^\mr{NLO}$}
\put(3,1){$=$}
\put(4,0.7){\line(1,0){3}}
\put(5.5,0.7){\circle*{0.2}}
\put(5.5,1.46){\circle{1.5}}
\put(7.5,1){$+$}
\put(8.5,1.14){\line(1,0){2.5}}
\linethickness{3mm}
\put(9.5,1.14){\line(1,0){0.5}}
\end{picture}
\end{center}
\vspace{-16mm}
\caption{\small Contributions to the pion self-energy at NLO: 1-loop graph with
a vertex from $L^{(2)}$ (little dot) and a counterterm from $L^{(4)}$ (fat
box). The divergent parts ($\propto \ep^{-1}$) compensate each other, and in
the finite parts ($\propto \ep^{0}$) the $\mu$-dependence cancels exactly.}
\end{figure}

What we are interested in is the prediction for the squared pseudo-Goldstone
boson mass in 2-flavour QCD versus the sum of the degenerate valence quark
masses, i.e.\ $\Mpi^2$ versus $2m$.
For the latter, we adopt ($\MSbar, 2\GeV$) conventions, which implies that the
leading-order (LO) constant $B$ is also quoted in $\MSbar$ scheme at
$\mu\!\sim\!2\GeV$, since the product $mB$ is scheme- and scale-independent
(as is $F$).
An important simplification stems from the fact that the GL-coefficients
$l_i^\mr{r}(\mu)$ may be traded for the low-energy scales $\Lambda_i$, and in
%terms of these the XPT predictions at next-to-leading (NLO) \cite{Gasser:1983yg}
%and next-to-next-to-leading (NNLO) \cite{Colangelo:1995jm,Burgi:1996qi} order
%take a form in which the scale-independence of the physical result is explicit
terms of these also the prediction at next-to-leading order (NLO)
\cite{Gasser:1983yg} in XPT takes a form in which the scale-independence of the
physical result is explicit (see \cite{Leutwyler:2000hx,Durr:2002zx} for
details and references):
\bea
\Mpi^2\!&\!{\mbox{\tiny LO}\atop\mbox{=}}\!&\!2mB\equiv:M^2
\label{Mpi0}
\\
\Mpi^2\!&\!{\mbox{\tiny NLO}\atop\mbox{=}}\!&\!M^2\:
\Big(1\!-{M^2\ovr32\pi^2F^2}\log({\Lambda_3^2\ovr M^2})\Big)
\label{Mpi1}
%\\
%\Mpi^2\!&\!{\mbox{\tiny NNLO}\atop\mbox{=}}\!&\!M^2\:
%\Big(1\!-{M^2\ovr32\pi^2F^2}\log({\Lambda_3^2\ovr M^2})+
%\label{Mpi2}
%\\
%&{}&\qquad{M^4\ovr256\pi^4F^4}
%\big[{17\ovr8}\log^2({\Lambda_M^2\ovr M^2})\!+\!k_M\big]\Big)
%\nonumber
%\;.
\eea

%%%%%%%%%%%%%%%%%%%%%%%%%%%%%%%%%%%%%%%%%%%%%%%%%%%%%%%%%%%%%%%%%%%%%%%%%%%%%%%

\section{$\!$AIMING AT NLO GL COEFFICIENTS}

\begin{figure}[!b]%3
\vspace{-8mm}
\epsfig{file=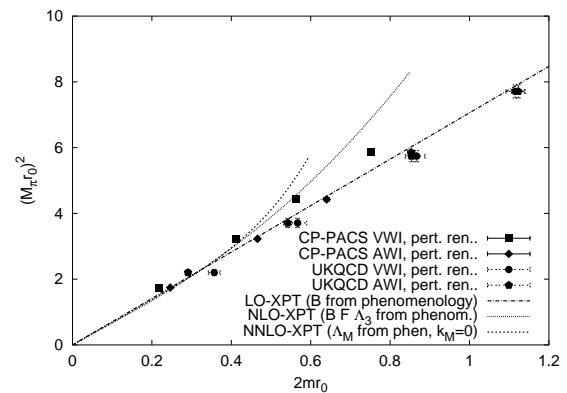,width=7.5cm,angle=0}
\vspace{-14mm}
\caption{\small LO/NLO/NNLO chiral predictions for $\Mpi^2$ (with
phenomenological values for the low-energy constants) compared to CP-PACS and
UKQCD data, using the VWI and AWI definition of the quark mass.}
%Note that the lines represent predictions, not fits.}
\end{figure}

\begin{figure}[!t]%4
\epsfig{file=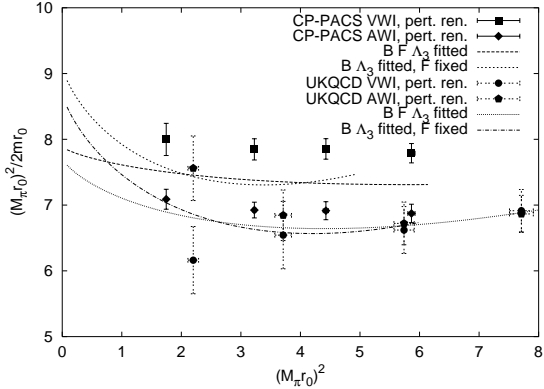,width=7.5cm,angle=0}
\vspace{-14mm}
\caption{\small Fits of the NLO functional form (\ref{Mpifit}) to the CP-PACS
\cite{AliKhan:2001tx} and UKQCD \cite{Allton:2001sk} data, after dividing by
$2m$. Either $F$ is kept fixed with the 3 most chiral points used or
$B, F, \Lambda_3$ are fitted, using all 4 points.}
\vspace{-6mm}
\end{figure}

%Equations (\ref{Mpi0}, \ref{Mpi1}, \ref{Mpi2}) may be used in two ways:
Equations (\ref{Mpi0}, \ref{Mpi1}) may be used in two ways:
Either phenomenological values for $B, F, \Lambda_3$ are {\em plugged in\/} and
the resulting curves are compared to lattice data, or the functional forms are
{\em fitted\/} to the data, which amounts to a lattice determination of the QCD
low-energy constants.

The first option is shown in Fig.\,3, where quark masses determined via the
axial (AWI) or vector (VWI) Ward-Takahashi identity
\cite{AliKhan:2001tx,Allton:2001sk} are renormalized by 1-loop perturbation
theory \cite{AliKhan:2001tx,Durr:2002zx}.
The heaviest pion is $\sim\!\!1.1\GeV$, and one sees that it is still
compatible with the LO graph, but not with the NLO or NNLO version.
The reason behind is that XPT yields an {\em asymptotic series\/}: the LO
result for $\Mpi$ may be good up to $\sim\!\!1\GeV$, while the NLO/NNLO forms
are more precise at small quark masses, but break down earlier, e.g.\ the NLO
functional form might be good up to (say) $600\MeV$, to NNLO form up to (say)
$400\MeV$.

\begin{table}[!b]%1
\vspace{-4mm}
\begin{flushright}
\begin{tabular}{|l|ccc|}
\hline
{} & $Br_0$ & $F_\pi r_0$ & $\Lambda_3 r_0$
\\
\hline
``phen.'' \cite{Gasser:1983yg,AliKhan:2001tx} & 7.06 & 0.233 & 1.51
\\
\hline
CP-PACS & 9.10 & (0.233) &  3.03
\\
CP-PACS & 7.88 &  0.503  &  3.97
\\
\hline
UKQCD   & 8.69 & (0.233) &  3.38
\\
UKQCD   & 7.70 &  0.318  &  3.46
\\
\hline
\end{tabular}
\vspace{-6.3mm}
\end{flushright}
\caption{\small Coefficients in the fits of the functional form
$(\ref{Mpifit})$ to the degenerate CP-PACS \cite{AliKhan:2001tx} and UKQCD
\cite{Allton:2001sk} data with average VWI and AWI masses \cite{Durr:2002zx}.
Constrained values in brackets, ranges such that $\#(\mr{d.o.f.})\!=1$.}
\end{table}

The second option is likely premature: If the estimate that the NLO functional
form is good up to $(\Mpi r_0)^2\!\sim\!2$ is correct \cite{Durr:2002zx} (with
$r_0\!\simeq\!0.5\fm$), then Fig.\,1 depicts the current situation, and there
is no room for using eqn.\,(\ref{Mpi1}) to fit the data, since it has 3
parameters and in either set at best 1 data-point is in the permissible range.
Still, one might simply go ahead and see how inconsistent results get, if
one ignores that the bulk of the data is likely not in a regime where (PQ-)XPT
applies.

For practical reasons, a factor $2m$ is taken out and parameters are made
dimensionless with powers of $r_0$, i.e.\ the actual NLO form used is
\beq
{(\Mpi r_0)^2\ovr 2mr_0}=
Br_0-{(\Mpi r_0)^2Br_0\ovr 32\pi^2(\Fpi r_0)^2}
\log{(\Lambda_3 r_0)^2\ovr (\Mpi r_0)^2}
\label{Mpifit}
\eeq
with the pertinent fits shown in Fig.\,4 and the resulting parameters
summarized in Tab.\,1 \cite{Durr:2002zx}.

Clearly, results should be taken with care, but it is reassuring to see that
the agreement among the fitted values of the NLO GL constant $\Lambda_3$ is not
substantially worse than for the LO constant $B$.
The numerical values in the last column of Tab.\,1 correspond to
$\Lambda_3\!\sim\!1.4\GeV$. This is consistent with the early estimate
$\Lambda_3\!\sim\!0.6\pm{1.4\atop0.4}\GeV$ \cite{Gasser:1983yg}, and hence
supports the hypothesis that at least in two-flavour QCD the {\em chiral
condensate is large\/} (see \cite{Leutwyler:2000hx} for references to this
topic).
However, the difference between $\Fpi$ from the unconstrained fit and the
physical value (cf.\ the central column in Tab.\,1) reminds one that to date
the bulk of the data is likely outside the regime of validity of (PQ-)XPT and
accurate simulation results at smaller quark masses are needed (cf.\
\cite{Farchioni:2002ca}) before one may draw definite conclusions.

\bigskip

In summary, $\Mpi^2$ vs.\ $2m$ illustrates that lattice QCD is, in principle,
an ideal tool to determine certain NLO GL coefficients.
To date, most studies with $\Nf\!\!=\!\!2$ Wilson quarks are likely in a
mass-range where there is only a ``point-like'' overlap with the regime where
(PQ-)XPT holds.
Once moderately lower masses are covered and with further improvements in
renormalizing the bare data and controlling the lattice artefacts (see
\cite{Durr:2002zx} for a discussion) the lattice will yield a competitive
determination of those NLO GL coefficients which encode how low-energy QCD
Green's functions depend on the quark mass.

%%%%%%%%%%%%%%%%%%%%%%%%%%%%%%%%%%%%%%%%%%%%%%%%%%%%%%%%%%%%%%%%%%%%%%%%%%%%%%%

\end{document}